\documentclass[aps,prl,twocolumn,showpacs]{revtex4}
\usepackage{amsmath}
\usepackage{epsfig}
\usepackage{amssymb}
\usepackage{dcolumn}
\usepackage{graphicx}
\usepackage{dcolumn}

\begin{document}
\date{\today}

\title{Lie symmetries and solitons in nonlinear systems with  spatially inhomogeneous nonlinearities}

\author{Juan Belmonte-Beitia}
\email{juan.belmonte@uclm.es}

\author{V\'{\i}ctor  M. P\'erez-Garc\'{\i}a}
\email{victor.perezgarcia@uclm.es}
\homepage{http://matematicas.uclm.es/nlwaves}

\author{Vadym Vekslerchik}
\email{vadym@ind-cr.uclm.es}

\affiliation{Departamento de Matem\'aticas, Escuela T\'ecnica
Superior de Ingenieros Industriales, and Instituto de Matem\'atica Aplicada a la Ciencia y la Ingenier\'{\i}a (IMACI),
Universidad de Castilla-La Mancha, 13071 Ciudad Real, Spain.}

\author{Pedro J. Torres}
\email{ptorres@ugr.es}

\affiliation{Departamento
de Matem\'atica Aplicada, Universidad de Granada, 
Campus de Fuentenueva s/n, 18071 Granada, Spain.}

\begin{abstract}
Using Lie group theory and canonical transformations we construct explicit solutions of  nonlinear Schr\"odinger equations with  spatially inhomogeneous nonlinearities. We present the general theory, use it to  show that localized nonlinearities can support bound states with an arbitrary number solitons and discuss other applications of interest to the field of nonlinear matter waves.
\end{abstract}


\maketitle

 \emph{Introduction.-}  Solitons are self-localized nonlinear waves which are sustained by an equilibrium between 
 dispersion and nonlinearity and appear in a great variety of physical contexts  \cite{SoliGen}. In particular, these nonlinear structures have been generated recently in ultracold atomic bosonic gases cooled down below the Bose-Einstein transition temperature \cite{dark,bright,gap}. In those systems the effective nonlinear interactions are a result of the elastic two-body collisions between the condensed atoms. 
 
 These interactions can be controlled by the so-called Feschbach resonance (FR) management \cite{FB1}, which has been used to generate bright solitons \cite{bright,Cornish}, induce collapse \cite{collapse}, etc. Recently, the control in time of the condensate scattering length has been the basis for many theoretical proposals to obtain different types of nonlinear structures such as periodic waves \cite{Kono1}, shock waves \cite{Kono2}, stabilized solitons \cite{Ueda}, etc.
 
 Interactions can  also be made spatially dependent by acting on the spatial dependence of either the magnetic field or the laser intensity (in the case of optical control of FR \cite{FB2}) acting on the Feschbach resonances. This possibility has motivated in the last years a strong theoretical interest on nonlinear phenomena in Bose-Eintein condensates (BECs) with spatially inhomogeneous interactions. Several phenomena have been studied in quasi-one dimensional scenarios such as the emission of solitons  \cite{Victor1} and the dynamics of solitons when the space modulation of the nonlinearity is a random \cite{Garnier}, linear   \cite{Panos}, periodic \cite{Boris2},  or localized function \cite{Primatarowa}. The existence and stability of solutions has been studied in  Ref. \cite{tuti}.  
 
  In this paper we construct general classes of nonlinearity modulations and external potentials for which explicit solutions can be constructed. To do so we will use Lie group theory and canonical tranformations connecting problems with inhomogeneous nonlinearities with simpler ones having an homogeneous nonlinearity.  We will show that localized nonlinearities can support bound states with an arbitrary number solitons without any additional external potential, something which differs drastically from the case of inhomogeneous nonlinearities. Our focus will be on applications to matter waves in BEC problems but our ideas can also be useful in the field of nonlinear optical systems \cite{nlop}.
 
 \emph{General theory.-} In this paper we will consider physical systems ruled by the 
 nonlinear Schr\"odinger equation with a spatially inhomogeneous nonlinearity (INLSE), i.e. 
\begin{equation}
 \label{NLSEin}
 i\psi_{t} = -\psi_{xx}+V(x)\psi+g(x)\left\lvert\psi\right\rvert^{2}\psi,
 \end{equation}
 where $V(x)$ is an external potential and $g(x)$ describes the spatial modulation of the nonlinearity. 
 Stationary solutions of the INLSE are of the form $\psi=\phi e^{-i\lambda t}$ where 
  \begin{equation}
 \label{estacionario}
 - \phi_{xx}+V(x)\phi+g(x)\phi^{3}=\lambda \phi, \quad\phi(\pm\infty)=0,
 \end{equation}
A second-order differential equation $A(x,u,u',u'')=0, $
possesses a Lie point symmetry  \cite{Bluman} of the form
$M=\xi(x,u) \partial/\partial x+\eta(x,u)\partial/\partial u$ if 
\begin{multline}
M^{(2)} A(x,u,u',u'') = \left[\xi(x,u)\frac{\partial}{\partial x}+\eta(x,u)\frac{\partial}{\partial u}+ \right. \\ \left. \eta_{1}(x,u)\frac{\partial}{\partial u'}+
\eta_{2}(x,u)\frac{\partial}{\partial u''}\right] A(x,u,u',u'') = 0.
\end{multline}
In our case, $A(x,\phi,\phi_{x},\phi_{xx})$ is given by Eq. \eqref{estacionario}
and the action of the operator $M^{(2)}$ on it leads to 
\begin{subequations}
\begin{eqnarray}
\label{relaciones}
\xi_{uu}&=&0,
\\
\eta_{uu}-2\xi_{ux}&=&0,
\\
2\eta_{xu}-\xi_{xx}-3f\xi_{u}&=&0,
\\
\eta_{xx}+\xi f_{x}+\eta f_{u}-\eta_{u}f+2\xi_{x}f&=&0.
\end{eqnarray}
\end{subequations}
Solving the previous equations, we find that the only Lie point symmetries of Eq. (\ref{estacionario}) are of the form
\begin{equation}
\label{simetria}
M=b(x)\frac{\partial}{\partial x}+c(x)\phi\frac{\partial}{\partial \phi},
\end{equation}
where
\begin{subequations}
\label{relaciones}
\begin{eqnarray}
g(x) & = & g_{0}b^{-3}e^{-2K\int_{0}^{x}1/b(s)ds},\\
c(x) & = & \tfrac{1}{2}b'(x)+K, \label{relacionesa}\\
0 & = & c''(x)-b(x)V'(x)-2b'(x) \left(V(x)-\lambda\right).\label{relacionesb}
\end{eqnarray}
\end{subequations}
for any constant $K$. Eqs. (\ref{relaciones}) allow us to construct pairs $\{ V(x), g(x) \}$ for which a Lie point symmetry exists.

\emph{Conservation laws and canonical transformations.-} It is known \cite{Leach}, that the invariance of the energy is associated to the translational invariance.  The generator of such a transformation is of the form $M=\partial/\partial x$. To use this fact, we define the transformation
\begin{equation}
\label{transformaciones}
X=f(x),\qquad U=n(x)\phi
\end{equation}
where $f(x)$ and $n(x)$ will be determined by requiring 
 that a conservation law of energy type $M=\partial/\partial X$ exists 
in the canonical variables. Using Eqs. (\ref{simetria}) and  \eqref{transformaciones}, we get
\begin{subequations}
\begin{eqnarray}
f(x) & = & \int_{0}^{x} \frac{1}{b(s)}ds,\\
 n(x) & = & b(x)^{-1/2}e^{-K\int_{0}^{x}1/b(s)ds}.
\end{eqnarray}
\end{subequations}
When $K=0$ the transformations preserve the Hamiltonian structure, and Eq.  (\ref{estacionario}) in terms of
$U=b^{-1/2}(x)$ and 
$X=\int_{0}^{x}1/b(s) ds$ becomes
\begin{equation}
\label{homogenea}
-\frac{d^{2}U}{dX^{2}}+g_{0}U^3=EU,
\end{equation}
where $E= \left(\lambda-V(x)\right) b(x)^{2}-\tfrac{1}{4}b'(x)^{2}+ \tfrac{1}{2}b(x)b''(x)$
 is a constant. This means that in the \emph{new variables we obtain the nonlinear Schr\"odinger equation (NLSE) without external potential and with an homogeneous nonlinearity}.  Of course not any choice of $V(x)$ and $g(x)$ leads to the existence of a Lie symmetry or an appropriate canonical transformation (e.g. the function $b(x)$ must be sign definite for $U$ and $X$ to be properly defined). 

\emph{Connection between the NLSE and INLSE via the LSE.-} We can use all the known solutions of the NLSE \eqref{homogenea}, e.g. solitons, plane waves and cnoidal waves  to construct solutions to Eq. (\ref{estacionario}).   Setting $K=0$ and eliminating $c(x)$ in Eqs.  \eqref{relaciones} we get 
  \begin{equation}\label{X1}
g(x)= g_{0}/b(x)^{3},
 \end{equation}
 and an equation relating $b(x)$ and $V(x)$
 \begin{equation}\label{X2}
 b'''(x)-2b(x)V'(x)+4b'(x) \lambda -4b'(x)V(x) =0. 
 \end{equation}
Although we can eliminate $b(x)$ and obtain a nonlinear equation for the pairs $g(x)$ and $V(x)$ for which there is a Lie symmetry,  it is more convenient  to work with \eqref{X2}, which is a linear equation.
Alternatively, we can define $\rho(x) = b^{1/2}(x)$ and get an Ermakov-Pinney equation \cite{both}
 \begin{equation}\label{Ek}
 \rho_{xx} + \left( \lambda - V(x)\right)\rho = E/\rho^3.
 \end{equation}
whose solutions can be constructed as 
 \begin{equation}\label{ep}
 \rho = \left(\alpha \varphi_1^2 + 2\beta \varphi_1\varphi_2 + \gamma \varphi_2^2\right)^{1/2},
 \end{equation}
 with $\alpha, \beta, \gamma$ constant and $\varphi_j(x)$ being two linearly independent solutions of the Schr\"odinger equation
 \begin{equation}\label{final}
 -\varphi_{xx} +  V(x) \varphi = \lambda \varphi.
 \end{equation}
This choice leads to $E = \Delta W^2$ with $\Delta = \alpha \gamma - \beta^2$ and $W$ being the (constant) Wronskian $W = \varphi'_1\varphi_2 - \varphi_1 \varphi'_2$. Thus, given any arbitrary solution of the \emph{linear Schr\"odinger equation  \eqref{final} we can construct solutions of the nonlinear spatially inhomogeneous problem Eq. (\ref{estacionario}) from the known solutions of Eq. (\ref{homogenea})}. Thus, using the huge amount of knowledge on the linear 
Schr\"odinger equation we can get potentials $V(x)$ for which $\varphi_1$ and $\varphi_2$ are known and 
construct $b(x)$, the canonical transformations $f(x),n(x)$, the nonlinearity $g(x)$ and the explicit solutions $\phi(x)$.

  \begin{figure}
 \epsfig{file=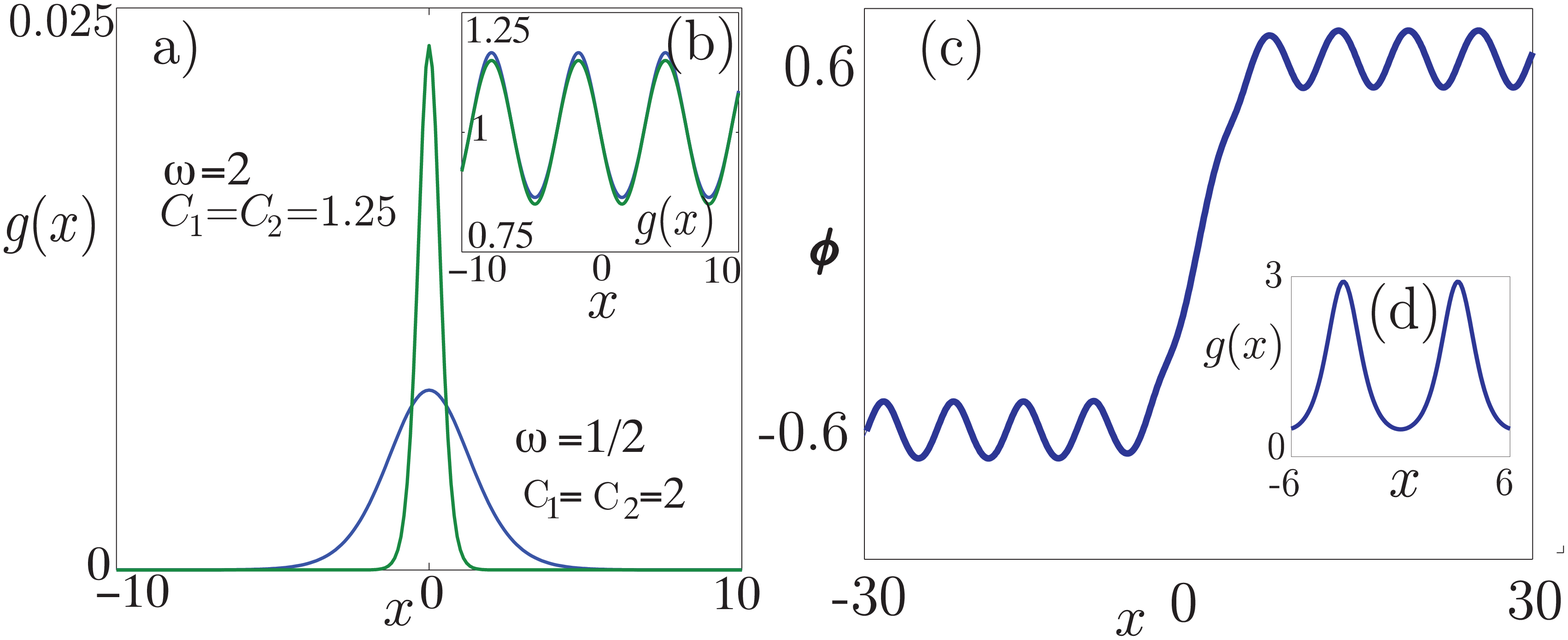,width=\columnwidth}
 \caption{[Color online] (a) Examples of exponentially localized nonlinearities given by  Eq. (\ref{b1a})with $C_1=C_2=2,C_3=0,\omega = 1/2$ (blue line), and
 $C_1= C_2 = 1.25, C_3=0, \omega = 2$ (green line). (b) Comparison of $g(x)$  
 given by Eq. (\ref{qper}) for $g_0=\omega=1, \alpha = 0.05$ (blue line) and its harmonic approximation (\ref{Tay}) (green line). (c) Example of black soliton solution of Eq. (\ref{estacionario}) with $V=0$ and an inhomogeneous nonlinearity given by Eq. (\eqref{qper}) with $\omega = 1, \alpha = 0.3$ and $g_0=1$. (d) Inhomogeneous nonlinearity given by Eq. \eqref{qper} with $\omega = 1, \alpha = 0.3$ and $g_0=1$ used to calculate the black soliton solution.
 \label{prima}}
\end{figure}
 % \begin{figure}
  % \epsfig{file=figdarksol.eps,width=5 cm}
 %\caption{[Color online] Example of black soliton solution of Eq. (\ref{estacionario}) with $V=0$ and an inhomogeneous nonlinearity given by Eq. (\eqref{qper}) with $\omega = 1, \alpha = 0.3$ and $g_0=1$.  \label{dual}}
%\end{figure}
 
\emph{Systems without external potential ($V(x) = 0$).-}
As a first application of our ideas let us choose $V(x)=0$, then 
Eq. \eqref{X2} becomes
$ b'''(x)+4b'(x)\lambda=0$ 
and its solution is
\begin{subequations}
\begin{eqnarray}
b(x) & = & C_1 \sin \omega x  + C_2 \cos \omega x  + C_3 \ \ ( \lambda > 0), \label{b1b} \\
b(x) & = & C_1 e^{ \omega x} + C_2 e^{-\omega x} + C_3 \ \ \ (\lambda <0), \label{b1a}
 \end{eqnarray} 
  \end{subequations}
 where $\omega = 2 \sqrt{|\lambda|}$. 
Using Eq. (\ref{X1}) we see that Eq. (\ref{b1a}) corresponds to an 
exponentially localized nonlinearity [Fig. \ref{prima}(a)] and 
Eq. (\ref{b1b}) leads to a periodic one
 \begin{equation}\label{qper}
 g(x) =  g_0 \left(1 + \alpha \cos \omega x \right)^{-3},
  \end{equation}
[Fig. \ref{prima}(b,d)]. For small $\alpha$ this nonlinearity is approximately harmonic [Fig.  \ref{prima}(b)]
 \begin{equation}\label{Tay}
 g(x) \simeq g_0 \left(1 - 3 \alpha \cos \omega x \right), \ \alpha \ll 1,
 \end{equation}
 %(see e.g. Fig. \ref{prima}(c)). 
We can construct our canonical transformation by using 
Eqs. (\ref{transformaciones}) and obtain 
 \begin{equation}\label{cuci}
  \tan \left( \frac{ \omega }{ 2 } \sqrt{1-\alpha^2} \; X(x) \right) = 
  \sqrt{ \frac{1-\alpha}{1+\alpha} } \; \tan \frac{\omega x}{2}.
\end{equation}
Using any solution of Eq. (\ref{homogenea}) 
with $E=\tfrac{1}{4}\omega^{2}\left(1-\alpha^2\right)$
this transformation provides 
solutions of Eq. (\ref{estacionario}) with $g(x)$ given by \eqref{qper}. 
For example, when $g_{0}>0$ we can obtain  black soliton 
solutions of Eq. (\ref{estacionario}) of the form
 \begin{multline}\label{brito}
 \phi(x)  =  
 \frac{\omega}{2}
 \sqrt{ \frac{1-\alpha^2}{g_0} 
 \left(1 + \alpha \cos \omega x \right)} \\ 
\times  \tanh\left[ 
    \frac{\omega}{2} \sqrt{ \frac{1-\alpha^2}{2} } \; X(x) 
 \right],
 \end{multline}
where $X(x)$ is defined by Eq. \eqref{cuci} [Fig. \ref{prima} (c)]. 
We emphasize that this is only a simple example of the many posible solutions 
that can be constructed in such a way.

\begin{figure}
 \epsfig{file=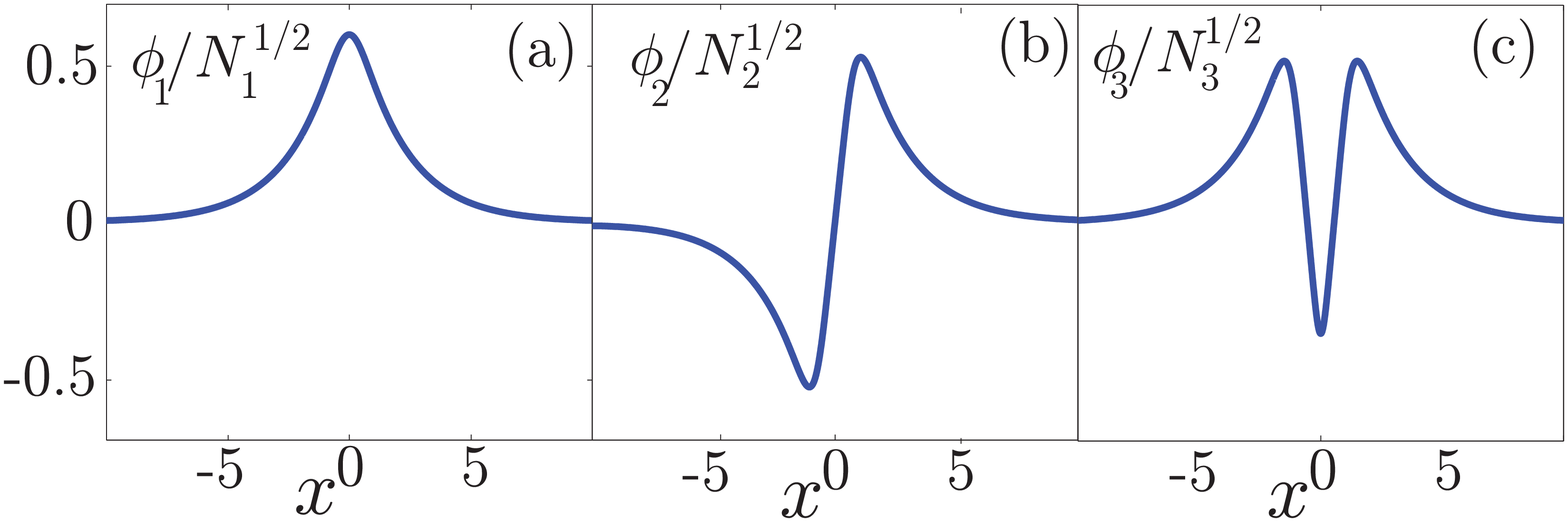,width=\columnwidth}
 \caption{[Color online] Solutions of Eq. (\ref{b1a-old-eq}) with $g_0=1$ corresponding to (a) $n=1, k_1 = 0.634493$ (b) $ n=2, k_2 = 0.690727$ and (c) $n=3, k_3 = 0.699957$. \label{elliptic}}
 \end{figure}

Concerning the case given by Eq. (\ref{b1a}), we would like to discuss it in more detail since we will get 
 an interesting phenomenon from its analysis. In order to simplify the following formulae (without loosing any significant features) 
we restrict ourselves to a particular choice of the constants $\Omega =1, C_{1}=C_2=1/2$ and $C_3=0$ in Eq. (\ref{b1a}), thus $ b(x) = \cosh x$
and Eq. (\ref{estacionario}) with $g(x)$ given by Eq. (\ref{X1}) and $\lambda=-1/4$,
\begin{equation}
  - \phi_{xx} + \frac{1}{4} \phi + \frac{g_{0}}{\cosh^{3}x} \phi^{3} = 0
\label{b1a-old-eq}
\end{equation}
in terms of $U$ and $X$ can be written as Eq. (\ref{homogenea}) with $E=-1/4$ being  $\cos X(x) = - \tanh x$, thus $0 \le X \le \pi$, and to meet the boundary conditions $\phi(\pm\infty)=0$ one has to impose
$U(0) = U(\pi) = 0$. This means that the original infinite domain in Eq. (\ref{b1a-old-eq}) is mapped into a bounded domain for Eq. (\ref{homogenea}). It is easy to check that when $g_{0}<0,$
\begin{equation}\label{ellip}
  U(X) = \eta \; \frac{ \mathop{\mbox{sn}}(\mu X, k) }{ \mathop{\mbox{dn}}(\mu X, k) }
\end{equation}
solves Eq. (\ref{homogenea}) provided
$ \mu^{2} = 1/\left[4 \left( 1 - 2k^{2} \right)\right]$ and $\eta^{2} = k^{2} \left( 1-k^{2} \right)/\left[2 \left|g_{0}\right| \left( 1 - 2k^{2} \right)\right]$.
The function $U(X)$ satisfies $U(0)=0$ and in order to meet $U(\pi)=0$, the condition $\mu\pi = 2nK(k)$ where $K(k)$ is the elliptic integral $ K(k) = 
  \int_{0}^{\pi} 
\left(1 - k^{2} \sin^{2}\varphi\right)^{-1/2} d\varphi$ must hold. 
Thus, to satisfy the boundary conditions, $k$ 
must choosen to satisfy 
$
4n K\left( k \right) 
  \sqrt{ 1 - 2k^{2} }
= \pi$ for $ n=1, 2, ...$. It can be shown that for every integer number $n$ this algebraic equation has only a solution $k_n$ which means that 
there are an infinite number of solutions of Eq. (\ref{b1a-old-eq}) of the form given by Eq. (\ref{ellip}). Moreover, each of those solutions has exactly $n-1$ zeroes. In Fig. \ref{elliptic} we plot some of them corresponding to $n=1,2,3$. These solutions can be seen as ``bound states" of several ($n$) solitons with alternating phases and their existence is remarkable. When  the nonlinearity is homogeneous, $g(x) = g_0<0$, Eq. (\ref{estacionario}) has only one localized solution for each $\lambda$, the cosh-type soliton, in other words: there are no bound states of several solitons. However, when $g(x)$ is modulated and decays exponentially 
as given in Eq. (\ref{b1a-old-eq}) \emph{we get an infinite number 
of localized solutions labelled by their finite number of nodes}. This is a novel and interesting feature of localized nonlinearities.

\emph{Systems with quadratic potentials $V(x) = x^2$.-} Any potential for which explicit solutions of Eqs. (\ref{final}) are known can be used to find nontrivial nonlinearities for which solutions can be constructed. Out of many possibilities we discuss only an example of interest for the applications to nonlinear matter waves in Bose-Einstein condensates which is $V(x) \propto x^2$. 

\begin{figure}
 \epsfig{file=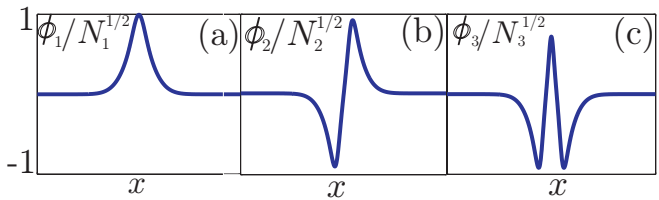,width=\columnwidth}
 \caption{[Color online] Different solutions of Eq. (\ref{estacionario}) for potential and nonlinearities given by Eq. (\ref{exa}). (a) $n=1$, (b) $n=2$  and (c) $n=3$. The spatial region shown corresponds to $x\in [-6,6]$. \label{ex2-fig}}
 \end{figure}

Let us choose $ b(x) = e^{ x^{2} }$, which leads to 
a quadratic trapping potential $V(x) = x^{2}$ and a gaussian nonlinearity such as the one generated by controlling the Feschbach resonances optically using a Gaussian beam  (see e.g. \cite{Victor2}), thus
\begin{equation}\label{exa}
g(x) = g_{0} \exp\left( -3x^{2} \right), \ \ \ V(x) = x^2.
\end{equation}
Our canonical transformation is given by 
$ X(x) =  \int_{0}^{x} dt \; \exp\left(-t^{2}\right) = \left(\sqrt{\pi}/2\right)  \mathop{\mbox{erf}} x$. 
In this case Eq. (\ref{estacionario}) is transformed into 
\begin{equation}
  - U_{XX} + g_{0} U^{3} = 0.
\label{ex2-new-eq}
\end{equation}
Note that the range of $X$ is again finite since
$ -\sqrt{\pi}/2 \le X \le \sqrt{\pi}/2$, 
and hence, we can again construct many \emph{localized} solutions to Eq.  
(\ref{estacionario}) starting from solutions of \eqref{ex2-new-eq} which satisfy the 
boundary conditions  $U( \pm\sqrt{\pi}/2) = 0$.
This can be done noting that for $g_{0}<0$ and any $\mu$ the functions 
\begin{equation}
  U^{(1)}(X) = 
  \frac{ \mu }{ \sqrt{\left| g_{0} \right|} } \;
  \mathop{\mbox{cn}}\left( \mu X, \; k_{*} \right) 
\end{equation}
and
\begin{equation}
  U^{(2)}(X) = 
  \frac{ \mu }{ \sqrt{2\left| g_{0} \right|} } \;
  \frac{\mathop{\mbox{sn}}\left( \mu X, \; k_{*} \right) }
       {\mathop{\mbox{dn}}\left( \mu X, \; k_{*} \right) }
\end{equation}
with $k_{*} = 1 / \sqrt{2}$
solve Eq.  (\ref{ex2-new-eq}) and that $U^{(1)}(X)$ and $U^{(2)}(X)$  vanish when 
$\mu X = (2n+1) K\left( k_{*} \right)$ and
$\mu X = 2n K\left( k_{*} \right)$ correspondingly. 
Thus we come to an infinite number 
of solutions of Eq. (\ref{ex2-new-eq}) under zero boundary conditions on the new finite interval, 
which correspond to different values of $\mu$.
Finally, localized solutions of the NLS equation  (\ref{estacionario}), are given by 
\begin{equation}
  \phi_{n}(x) = 
  \left\{
  \begin{array}{lcl}
    \frac{ 2n K\left( k_{*} \right) }{ \sqrt{ \pi\left| g_{0} \right| } } \;
    e^{x^{2}/2 } \;
    \mathop{\mbox{cn}}\left( \theta_{n}(x), k_{*} \right)
  && n=1,3,...
  \\
    \frac{ 2n K\left( k_{*} \right) }{ \sqrt{ 2\pi\left| g_{0} \right| } } \;
    e^{x^{2}/2} \;
    \frac{\textstyle \mathop{\mbox{sn}} \left( \theta_{n}(x), \, k_{*} \right) }
         {\textstyle \mathop{\mbox{dn}} \left( \theta_{n}(x), \, k_{*} \right) }
  && n=2,4,...
  \end{array}
  \right.
\label{ex2-phi}
\end{equation}
with
\begin{equation}
  \theta_{n}(x) = 
  n K\left( k_{*} \right) \mathop{\mbox{erf}} x
\end{equation}
It can be shown by simple asymptotic analysis that the last factors in Eq. (\ref{ex2-phi}) 
tend to zero as $x \to \pm\infty$ faster than $\exp( -x^{2}/2 )$ and that 
these are indeed \emph{localized} solutions of our problem 
as it can be seen in the ones plotted in Fig. \ref{ex2-fig}, with different numbers of zeroes 
($\phi_{n}(x)$ possesses $n-1$ zeroes).

In conclusion, we have used Lie symmetries and canonical transformations to construct explicit solutions of the nonlinear Schr\"odinger equation with a spatially inhomogeneous nonlinearity from those of the homogeneous nonlinear Schr\"odinger equation. The range of nonlinearities and potentials for which this can be done is very wide. We have studied in detail the case $V=0$ with localized and periodic nonlinearities. In the former case, we have used our theory to construct an infinite number of multi-soliton bound states, something which is not posible in the case of spatially homogeneous nonlinearities. Finally, we have presented an example of physical interest (harmonic trap and gaussian nonlinearity) for which exact solutions can be constructed. The ideas contained in this paper could also be applied to study time dependent problems, higher-dimensional situations, multi-component systems, etc. We hope that this paper will stimulate further research on those topics and help to understand the behavior of nonlinear waves in systems with spatially inhomogeneous nonlinearities.

\acknowledgements

This work has been partially supported by grants BFM2003-02832, FIS2006-04190, MTM2005-03483
 (Ministerio de Educaci\'on y Ciencia, Spain) and PAI-05-001 (Consejer\'{\i}a de Educaci\'on y Ciencia de la Junta de Comunidades de Castilla-La Mancha, Spain).

  \end{document}